# Designing Accessible Visual Programming Tools for Children with Autism Spectrum Condition

Misbahu S. Zubair[a,b,*], David J. Brown[a], Matthew Bates[a] and Thomas Hughes-Roberts[c]

[a]School of Science and Technology, Nottingham Trent University, Clifton Campus, NG11 8NS, Nottingham, UK.
[b]Software Engineering Department, Faculty of Computer Science and Information Technology, Bayero University, Kano, Nigeria.
[c]Department of Computer Science, Liverpool John Moores University, L3 3AF, Liverpool, UK.

## ABSTRACT

Visual Programming Tools (VPTs) allow users to create interactive media projects such as games and animations using visual representations of programming concepts. Although VPTs have been shown to have huge potential for teaching children with cognitive impairments including those with Autism Spectrum Condition (ASC), research has shown that existing VPTs may not be accessible to them. Therefore, this study proposes a set of recommendations for the design of accessible VPTs for children with ASC. Recommendations were initially gathered and validated by interviewing experts (n=7). The interviews were thematically analysed to identify recommen-dations. A second set of interviews with a subset of the initial experts (n=3) was then conducted to validate the gathered recommendations. An examination of the available literature was then conducted to identify additional recommendations for the design of VPTs. These recommen-dations arose from those used for the design of other interactive applications for children with ASC (e.g. virtual environments, serious games) and not identified as part of those the initially gathered from interviews. A novel set of recommendations for the design of VPTs for children with ASC and additional cognitive impairments has been defined as the result of this study.

## 1. Introduction

Autism Spectrum Conditions (ASC) sometimes referred to as Autism Spectrum Disorders (ASD), are diagnosed by the presence of social, communication and interaction difficulties, together with repetitive patterns of behaviour and restricted interests (American Psychiatric Association, 2013; Baron-Cohen et al., 2009). The individual conditions that now makeup ASC (e.g. autistic disorder, Asperger's disorder, and pervasive developmental disorder) used to be considered as distinct conditions or disorders, however, the latest Diagnostic and Statistical Manual of Mental Disorders (DSM-5) consolidated them into ASD (American Psychiatric Association, 2013). According to the new manual, the symptoms of these disorders "represent a single continuum of mild to severe social communication and restrictive repetitive behaviours/interests". Although ASC can exist with any level of intelligence, it is more common among individuals with a learning disability, so much that there has been a debate on whether the two conditions are distinct (O'Brien and Pearson, 2004). The United Kingdom's Department of Health even includes individuals with ASC as part of those with learning disabilities as long as they do not have average or above-average intelligence(Department of Health, 2001).

Technology-based interventions have long since been used for individuals with disabilities, including those with ASC. A major advantage offered by this approach compared to traditional approaches is flexibility, i.e. it can be easily customised to meet the varying needs of users, and adjusted automatically based on each user's interactions (Allen et al., 2016). The emergence of smartphones and tablets has led to increased interest in technology-mediated education and interventions for individuals with ASC. This is because, in addition to the advantages offered by previous technological platforms, the mobility of such mobile devices extends the learning environment from classrooms to homes, which promotes learning (Allen et al., 2016). Another important contributor to the popularity of technology-based intervention is the reported appeal they have to children with ASC compared to traditional intervention approaches (Allen et al., 2016; Goldsmith and LeBlanc, 2004).

---

*Corresponding author

✉ misbahu.zubair@ntu.ac.uk (M.S. Zubair); david.brown@ntu.ac.uk (D.J. Brown); matthew.bates@ntu.ac.uk (M. Bates); t.hughesroberts@ljmu.ac.uk (T. Hughes-Roberts)

Using these modern technological platforms, interactive applications have been utilised in providing interventions for children with ASC. They have been used for improving communication skills (Alessandrini et al., 2014; Chien et al., 2015; Lorah et al., 2015), social skills (Alessandrini et al., 2014) collaboration skills (Zancanaro et al., 2014), and academic skills (Knight et al., 2013).

A technology-based approach used to teach skills and educational content to typically developing children that is not being fully utilised for children with ASC is content creation using Visual Programming Tools (VPTs). Visual programming allows users "specify a program in a two (or more) dimensional fashion" (Myers, 1990). During the last two decades, VPTs have been created specifically for children to learn to programme and to create content related to their interests e.g. animations, interactive stories and games. They represent programming concepts as blocks or bricks that can be put together to create a program and include a huge library of media resources that children can use to create programs using character sprites or models, background images or landscapes, and sounds.

This paper presents a study built on a previous investigation into the accessibility of VPTs (Zubair et al., 2018). This study has gathered a set of recommendations to guide the creation of accessible, usable and engaging VPTs for children with ASC, specifically those that also have a learning disability. Recommendations were gathered and validated with experts by collecting data using interviews and subsequent thematic analysis of these data. An examination of the available literature was then conducted to identify additional recommendations for the design of VPTs. These recommendations arose from those used for the design of other interactive applications for children with ASC (e.g. virtual environments, serious games) and not identified as part of those the initially gathered from interviews. Throughout this paper, the target population of this research, children with ASC, are sometimes referred to as users, especially when presenting the gathered recommendations.

### 1.1. VPTs for Children

This section briefly introduces the main VPTs found in the literature and/or used in practice.

*Scratch* is the most popular VPT for children based on findings in the literature. Using Scratch, children aged between 8 – 16 years can learn to programme while creating "interactive, media-rich projects" such as animations or games (Maloney et al., 2010). Scratch has a rich media library for images and sounds, an inbuilt sound recorder, a painter for creating custom images, and allows the import of existing media files. Each Scratch project consists of 2D media files controlled by a script made up of colourful command blocks snapped together. It has a single user interface with multiple panes to ease navigation, provides an error-free programming environment i.e. there is no wrong way of snapping blocks that fit together, makes data concrete by displaying variables on-screen during execution, and visualises execution by highlighting blocks that are executing at run-time. Scratch is mainly accessed through web browsers and is not primarily designed for use on mobile devices.

*Pocket Code* is a Scratch inspired mobile VPT targeted at children for creating animations and programs while learning to programme. Scratch inspirations such as block-based programming metaphors and an online community can be found in Pocket Code which targets users between 13 – 18 years old (Koitz and Slany, 2014). However, unlike Scratch, Pocket Code is mobile-based which means it has access and can make use of mobile sensors.

*Alice* is an innovative three-dimensional VPT created by the Stage 3 Research Group at Carnegie Mellon University (Cooper et al., 2000). It allows programming novices to create virtual worlds with animations, interactive narratives, and games. Projects in Alice are made up of 3D objects, each object holds its private properties (e.g. height and width) and member methods. At the start of a project, objects are added to the virtual world and positioned based on the needs of the user. Once objects are initialised, the program code is added using a drag and drop smart editor.

*Kodu* is a VPT created specifically for young children to learn to programme through independent exploration (MacLaurin, 2011). It is built within a real-time 3D game and includes features to assist users in creating content such as a terrain editor, layout tools, character menus etc. It addresses complexity by predefining physics, collision detection and the use of a camera control. Kodu is packaged with several sample worlds that the user can edit, interactive lessons, and an online community for sharing worlds with other Kodu users. Programming in Kodu, similar to Scratch, is block-based, where blocks of instructions are put together using a 'when - do' approach. In the 'when' slot, sensors are declared to determine a condition, and the in the 'do' slot, the actions that should be performed are listed.

## 1.2. VPTs as Learning and Teaching Aids for Neurotypical Children

Due to the graphical representation of programming concepts provided by VPTs, the most obvious application for children is in the context of learning programming visually instead of using text-based programming languages. A study by Meerbaum-Salant et al. (2013) showed that middle school students with no previous programming experience achieved a reasonable understanding of computer science concepts using Scratch. The findings of a study by Fokides (2017) with 138 5th grade students using Kodu to create games also showed that students were able to learn basic game design and programming concepts as they authored their first games.

Furthermore, proof of the suitability of VPTs for learning programming has been found when comparing the level of students' programming abilities and misconceptions based on the programming language used to learn (Mladenović et al., 2018). More than 200 10 - 12-year-old students from four schools participated in this study and were divided into six subgroups. Every two subgroups were assigned a separate programming language to learn to programme with i.e. two subgroups used Logo, another two subgroups used Python, and the last two subgroups used Scratch. Pre and post-tests were used to assess the participants' understanding of sequencing and loops. The results showed that misconceptions about loops were minimized when visual block-based Scratch was used compared to when text-based Python and Logo were used.

Another context in which VPTs have been applied is in learning specific academic subjects. Ke (2014) implemented a design-based approach to learning mathematics with 64 middle school students. Participants worked within design groups and played several educational mathematics games before creating their own games using Scratch. The evaluation demonstrated improved attitudes towards mathematics by the students after participating in the game making process.

Storytelling is also a context in which VPTs are used. Burke and Kafai (2010) investigated the possibility of creating Scratch computer programs as a way of developing the creative and storytelling abilities of children. Middle school children aged between 11 - 14 years took part in the study and 60% of participants felt their storytelling skills improved.

In addition to learning programming, computational concepts, academic subjects, and improving creativity, additional benefits of using VPTs have been reported. For example Sáez-López et al. (2016) found that the use of VPTs also resulted in motivation, fun and enthusiasm on the part of the children.

## 1.3. VPTs as Learning and Teaching Aids for Children with ASC

Despite the numerous reports and studies in the literature supporting the advantages of using VPTs to teach various skills, few studies reported the use of such tools by children with ASC, and fewer reports still of use by children with ASC and associated learning disability.

To improve the problem solving and creative skills of children with high functioning ASC, Sarachan (2012) proposed a workshop where these children will use Scratch or similar tools to explore their interests in computers and games.

Computational thinking skills of children with ASC can also be improved by the use of VPTs as shown by the study conducted by Munoz et al. (2018). Their study involved seven children aged between 11 - 15 years participating in a five week workshop for creating games using Scratch. Results showed encouraging improvements in the computational skills of the participants, especially considering the fact that they had no prior programming knowledge.

Social skills of children with ASC could also be improved with the help of VPTs as reported by (Eiselt and Carter, 2018). They conducted an 8-week programming course with eight children with high functioning ASC aged between 9 – 15 years which aimed to teach programming and social skills through game making. By the end of the study, the authors observed improved knowledge in programming and social interactions in participants. It is however interesting to note the findings reported by Bossavit and Parsons (2017) from their study with two teenagers with ASC programming a game of their choice using Kodu. The findings report the demonstration of problem-solving and programming skills by the participants, but very little interaction and collaboration between the two participants.

Overall the literature yields few studies related to the use of VPTs by children with ASC. The few that exist also relate to individuals with no reported associated learning disability. Zubair et al. (2018) study on evaluating the accessibility of Scratch for children with cognitive impairments, observed seven participants with cognitive impairments (including five with ASC) using Scratch to create stories and interviewed their class teachers after these observations had taken place. Findings from the observations revealed that the participants struggled with a number of user interface related, and cognition related accessibility issues. Findings from interviews with teachers revealed that these issues were caused by the primary use of text labelling, Scratch's lack of constraints, and the unavailability of templates to scaffold and guide users.

Although Bossavit and Parsons (2017), Eiselt and Carter (2018), and Munoz et al. (2018) did not report any accessibility issues, very little is known about the cognitive abilities of the users that participated in these studies. It is possible that unlike the 5 participants with ASC that participated in the study conducted by Zubair et al. (2018), the participants in the three studies above do not have an associated learning disability. This means that they may have an average or above-average IQ, thus they may not face the same difficulties in using VPTs as those with ASC and related learning disabilities.

### 1.4. Aim of the Study

This research aims to gather and propose a set of recommendations for designing usable, accessible and engaging VPTs for children with ASC, including those with a learning disability. In doing so it is hoped that more children with ASC will be able to use VPTs, with potential benefits reported to include improved engagement and collaboration (Hughes-Roberts et al., 2019).

## 2. Gathering Recommendations

The following section describes the qualitative study designed to gather design recommendations for accessible VPTs for children with ASC.

### 2.1. Participants

Experts (teachers with experience in working with children with ASC in special needs schools, and academic researchers with research interests and published works in areas related to the characteristics and needs of children with ASC, or the use of technological interventions for children with ASC) were the main participants of this research. They were recruited through two approaches; from existing links with Special Education Needs (SEN) schools and academic researchers in the relevant fields; and by searching online for academic experts with relevant research experience. Seven experts (E1 - E7) agreed to participate (three males and four females; two teachers and five academic researchers). Their experiences are discussed in detail in Table 1.

Table 1: Interviewed experts, their professions and experience.

| Expert | Profession | Experience |
| --- | --- | --- |
| E1 | Researcher | Associate professor in education with research interests within the field of special education needs and disabilities including the use of technology to provide personalised learning experiences for those with disabilities. Also provides training to teachers, carers and parents. |
| E2 | Researcher | Senior researcher with more than 10 years research experience in the area of assistive technologies. Recent research interests include the use of virtual reality for children with ASC. |
| E3 | Teacher | Assistant head teacher at a school for students aged from 3 - 18 years with severe learning difficulties, profound and multiple learning difficulties and ASC. Also has more than 20 years experience in the SEN field. |
| E4 | Researcher | Professor in the field of learning disabilities with more than 30 years research experience in the field. Expertise in the design and evaluation of technology for students with a range of cognitive impairments and ASC including virtual environments, serious games and robotics. |

Table 1: Interviewed experts, their professions and experience.

| Expert | Profession | Experience |
|---|---|---|
| E5 | Researcher | Associate professor with research interests in the behaviour of individuals with ASC, the use of interactive applications (e.g. video games) as a method of intervention or to understand their behaviours. Has received research funding from several bodies from multiple countries, and has collaborations in several countries. |
| E6 | Researcher | Associate professor with research interests in autism in education. Over 20 years research experience in various university research centers in multiple countries. |
| E7 | Teacher | Classroom teacher with several years of experience at a school for students aged from 3 - 18 years with severe learning difficulties, profound and multiple learning difficulties and ASC. |

### 2.2. Method

All seven participants were interviewed between May and August 2018. Ethical approval was sought and received from Nottingham Trent University's Ethics Committee before participants were recruited and interviews were conducted. Steps were also taken to ensure anonymity and confidentiality of participants. Six of the interviews were conducted face to face, and one was conducted via video conferencing (Skype). All interviews lasted between 30 – 60 minutes and were audio-recorded with the permission of the interviewee.

Each interview started with the expert being shown an example of a VPT (Scratch), and consisted of two other main parts that utilised observational data of five children with ASC using Scratch to create stories gathered in a previous study (Zubair et al., 2018). The first part which focused on grouping the five children based on their observed behaviours and discussing their characteristics and goals was used to collect data for a study on constructing personae for children with ASC (Zubair et al., 2019). The second part collected data for this study, and was mainly driven by discussions around all observed behaviours especially those related to difficulties associated with using VPTs, and ways of designing accessible VPTs that could be used by children with ASC to achieve their goals with little or no difficulties.

Using a semi-structured approach, the researcher was flexible and opportunistic in following up on relevant and interesting points as they arose, and by using the points raised by experts to decide the sequence in which behaviours and difficulties were discussed. In some cases, experts suggested mitigating or avoidance approaches for difficulties and approaches for encouraging positive behaviours without being asked. In other cases, the researcher asked for recommendations by following up on a point made about the behaviour or difficulty.

### 2.3. Analysis

The interviews were first transcribed, then analysed using thematic analysis (Braun and Clarke, 2006). Specifically, theory-driven thematic analysis was used to analyse the interview transcripts. In this approach, the researcher codes and creates themes to answer specific questions that are being asked of the data (Braun and Clarke, 2006). Two questions were asked in this case, the first question asked, "What features should visual programming tools have to ensure that they are accessible, usable and engaging for children with ASC?" and the second asked, "What features should visual programming tools avoid to ensure that they are accessible, usable and engaging for children with ASC?".

In analysing the transcribed data, all excerpts of transcripts that directly referenced a recommendation for the inclusion or exclusion of a feature were coded, as were excerpts describing features/approaches that worked or did not work well for the same target group either applied to other interactive applications or in real life. This approach was helpful in highlighting recommendations explicitly suggested by experts, as well as justification for making recommendations based on insights provided by experts.

After coding the interview transcripts and tagging relevant excerpts of the data as codes, the first round of theme creation was performed by grouping together related codes to form potential themes. Themes in this context represent the area of concern associated with a recommendation or group of recommendations. Coded extracts for each theme were then checked for consistency with each other, and the theme. This led to the elimination of redundant extracts, coded extracts that did not fit the code or theme, and ultimately led to the creation of nine consistent and meaningful themes. A final pass at reviewing the themes, their codes and extracts was conducted. This time all codes and extracts were found to be consistent with their themes.

QSR Nvivo 12 was used to conduct this data analysis. It was used to transcribe audio interview data, store notes, code data, and store themes.

Each theme and its related recommendations are discussed next. Excerpts from the transcripts showing experts' recommendations related to each theme are provided to show the opinions of experts in their own words.

### 2.4. Findings

The nine themes identified from the transcripts cover areas of concerns ranging from the type of hardware platform most suitable for children with ASC, to the type of content that should be provided within a VPT and how it should be presented. For each theme discussed in this section, relevant excerpts from the transcripts are provided to highlight the opinions of experts in that area of concern. The complete list of gathered recommendations organised by themes and their descriptions is also provided in Table 2.

#### 2.4.1. Mobile Device Compatibility

The use of VPTs requires constant and continuous interaction with visual representations of programming logic (e.g. blocks and bricks), and objects that are being programmed (e.g. character and background images). These interactions, as pointed out by E6, may not be straightforward for all users with ASC, especially using a mouse as an input device.

*"They need to be learning the link between what they're doing... some of them play video games and they're very clear that what they do with their hands has an effect on what happens on the screen, but with other children, making that link, particularly if they're going to use a mouse can be difficult."* (E6)

A potential solution to this problem is making VPTs compatible with and accessible on mobile touch screen devices such as smartphones and tablets. This could greatly reduce the distress caused by difficulties in interacting with other platforms that offer less intuitive input techniques. In addition to the directness of interaction and intuitiveness offered by touch screen devices, E4 points out that most young people including those with ASC own and use smartphones and tablets, and are comfortable interacting with objects on a touch screen.

*"A lot of young people use iPads obviously, and they get used to controlling things with their finger. And that conceptual link between you touching it and it doing something is a much shorter jump isn't it, it's much easier to understand what's happening there. Whereas with the mouse when you have to do the double click on things, some of that will be difficult for some children. Even if you tell them they have to click twice, they leave it so long that it does not work as a double click."* (E4)

The need for an accessible interaction method cannot be overstated because as explained by E5, interaction difficulties can cause a ripple effect leading to difficulties in other areas.

*"For example, if you have difficulty performing mouse operations then maybe you're also going to have difficulty structuring and sequencing, not because you fundamentally have difficulty structuring and sequencing but because you can't click the mouse in order to tell the computer what to do."* (E5)

#### 2.4.2. Engaging Users

Individuals with ASC are known to have restrictive interests; these interests may seem random, peculiar and sometimes strange. However, experts believed some core areas of interests are common among individuals with ASC, and integrating content related to these areas of interest within VPTs could potentially lead to higher engagement levels and motivate users to create programs around their areas of interests. E3 and E6 had this to say on the subject:

*"I mean there are some things that sort of generally float their boats in terms of ASC. I'm sort of being very generalistic here, for example, Thomas the Tank Engine, you know, they like the repetition, they like that it's not*

*unpredictable and it's a very closed world."* (E3)

*"So there are people who have weird and wonderful interests but there are some things that a lot of young people are interested in and it's a bit different for boys and girls. So, for younger children, we know Thomas the Tank Engine is you know… they actually love Thomas. For older boys, a lot of boys still like trains but not Thomas. With girls, they tend to love animals, you know. There are probably a number of core things that you could perhaps say: 'we know that a lot of boys like these things and a lot of girls like these things and maybe we'll focus on integrating those'."* (E6)

In addition to proving content related to known core areas of interests of children with ASC, another suggestion made by E3 is to use content receiving high levels of interest from users within a VPT as a potential choice (to build projects around) for other users that are not motivated.

*"If a child is motivated by one character, maybe you could take that and get rid of the others and use it as a choice for somebody that is less motivated."* (E3)

*2.4.3. Information Presentation and Visualisation*

The use of visual icons and symbols instead of text, or to support text, was a key suggestion by experts. This is due to the reading and reading comprehension difficulties, and preference for visual means of communication associated with children with ASC. E6 expressed concern about having more text labels than visual labels in VPTs:

*"I mean there is the issue of having to be able to read for meaning, it's not just reading the words, it's reading and understanding what it is, you know. If I read that, what will I have to do? So you have to have a certain level of reading comprehension, don't you? I know there are other online programs that are directed at young people with a certain level of intellectual ability and reading ability. So if it's something that's supposed to be accessible to everyone perhaps there needs to be a bit less writing and more visual kind of representations."* (E6)

E4 also shared similar concerns about users needing to read and comprehend textual labels in order to use a VPT. To illustrate the difficulty of the task, E4 compared it to reading a foreign language that one cannot speak:

*"Think of it as if you were a native Italian speaker looking at this (scratch interface in English) and didn't know any English how much of it could you understand?"*

Therefore, as E6 mentioned, just like other applications targeted at children with disabilities in general, VPTs for children with ASC should present information visually, with as many visual symbols and icons as possible to make it easy for users to understand without the need to read text or text labels. E1 also made the same recommendation as seen in the excerpt below:

*"You could have a version that minimized the need to read or eradicated the need to read by putting symbols on it. You can have a run icon, you can have a jump icon and lots of other things."* (E1)

*2.4.4. Sounds*

ASC is commonly associated with high sensory sensitivities, and affected children can be highly sensitive to sounds. Therefore, E4 expressed concern over the use of sounds in VPTs:

*"Is auditory feedback necessary? Because I don't think at the moment that's a good idea, it may well be that the auditory system could be overloaded for some people with ASC."* (E4)

Since sounds may not cause distress for all children with ASC and could be useful in content creation and program execution, the recommendation proposed here is to make sounds optional for the use of VPTs and only played when turned on. Therefore, feedback should also be provided through other means e.g. visual feedback for those that are sensitive to sounds and therefore have sounds turned off.

### 2.4.5. Restrictions and Limitations

Experts unanimously believed that the numerous choices (e.g. sprites, backgrounds, sounds, blocks) and open-ended scenarios (e.g. blank projects) presented by VPTs can be overwhelming for children with ASC. This is true even in their everyday activities as explained by E3:

*"When you are working with autistic students and you ask them an open question they struggle with that. So for example, if I was to say 'how do you feel?' there are sort of many different answers you can give. But if I say 'do you feel happy or sad today?' it sort of closes it down."* (E3)

According to E4, another negative effect of having this level of choice comes in the form of children with ASC performing repetitive tasks as a way of controlling their arousal levels.

*"There's a lot of theory around arousal levels and emotional disengagement in autism. A lot of the things observed [by the researcher] are attempts by the person to control their level of arousal. You are confronting somebody with a lot of information [within the VPT], and their return to doing repetitive tasks or something else is their attempt in saying 'I can do this, let's just carry on doing this'."* (E4)

Therefore, experts suggested the use of restrictions or limitations (when and where appropriate) as a way of providing safe and manageable options for children with ASC. This could potentially lead to them being comfortable enough to try out new things without getting overwhelmed. E6 suggests how this approach can be applied to providing a manageable number of media elements (e.g. story characters) to use in projects within a VPT:

*"So, you wouldn't make 50 characters available, you maybe make five available, so they'd just pick from a really small number so that they're not getting overwhelmed by the sheer number and not knowing what to pick. If there's a small number, they can pick on what they like the look of and then quickly get into doing something with it."* (E6)

In addition to media elements such as characters, limitations or restrictions can also be applied to other aspects of VPTs such as tools, features, programming elements etc. However, E1 cautioned that there might be a downside to this as having too many restrictions may prevent engagement.

*"So, the educator in me thinks I wouldn't want the restrictions to be there all the time, because I would be wanting to support them to do something more, but I would also want to be having it open enough so that they can get engaged in the first place"* (E1)

Therefore, a balance should be struck where restrictions exist to prevent users from getting overwhelmed and reacting negatively, but not too much to prevent users from engaging with the VPT.

### 2.4.6. Scaffolding

Experts believed that children with ASC may have difficulties with creativity and imagination, sequencing events, understanding and using programming elements (even if they are visual). They recommended the use of scaffolding to guide the users through the various stages of program creation, from media selection to program logic specification. As E6 explains below, individuals that require scaffolding to perform tasks that require sequencing and structuring (e.g. writing a story) without the use of a VPT will surely require scaffolding from the VPT if they are to use it to perform the same task independently.

*"Sequencing and structuring may well be things that those children have difficulty with when they're writing a story by hand in a book, you know, some children have that difficulty. So again, there might be a need for a bit more scaffolding for children who might have that difficulty. If you just sit them in front of a piece of software and ask them to create a story, that will be really difficult."* (E6)

Children with ASC also find it difficult to visualise imaginary things, settings or situations as pointed out by E3:

*"A lot of students struggle with imagination. [Creating content using VPTs] is equivalent to say role-play, unless you've actually got something in front of them, some props, then they will struggle just to make a story out the sky."* (E3)

Scaffolds can also help children with ASC avoid this difficulty by showing them what a possible outcome can look like, hence making it easy for them to create.

*"Having a sense of direction in the sense of reward, not reward but a clear idea of what an outcome might look like... I think sometimes young learners in general, but especially autistic groups, if you show them what the end goal is going to look like, they can better see the journey. And I think if you can provide scaffolding, thus enabling better structure and sequencing and things like that. I think that will better enable them to stay on track."* (E2)

Another potential benefit of having a scaffold as seen in the extract above is keeping users on track to achieve their goal and preventing them from getting lost in unrelated tasks.

Based on insights gathered from experts, we recommend the use of templates that can guide users through program creation, as well as programming specific scaffolds through the provision of highly abstracted programming blocks that represent scripts for performing popular actions.

### 2.4.7. Goal Orientation

Findings show that experts believe designing VPTs with specific goals in mind can make it easier to identify the aspects of VPTs that need to be eliminated, restricted, automated or greatly simplified to allow children with ASC to focus on achieving goals and obtaining feedback. For example, E6 had this to say about eliminating reading activities to get users to focus on programming:

*"You want them to try and do something quickly you know, so you don't want them to have to read things and be concentrating, it's not a reading exercise it's a programming exercise. So if they're trying to read and work out what they have to do, it slows things down to the extent that they lose interest, and then you have lost purpose... If the purpose of the activity is to learn programming using the program then it needs to be easily accessible so they can do it quickly because then they'll get feedback, "wow I can make that work" and then they will want to do something more complicated."* (E6)

This approach could be a way of reducing frustration, increasing motivation and goal achievement, and encouraging actual visual programming. This, in turn, can lead to continued use of the VPT. However, success needs to be achieved quickly for this to occur as E4 mentions below:

*"If you don't understand [visual programming], more exposure is going to help you with that. But then you're not going to get more exposure unless you get some degree of success. One thing we tried to do when we were designing stuff was trying to make sure that people could actually get something done, there will always be some degree of challenge but they always got some degree of success because if they don't, they just give up."* (E4)

### 2.4.8. Personalisation

Children with ASC face varying degrees of difficulties, and experts stressed the need to acknowledge this when designing VPTs: *"you need some differentiation built-in"* (E2). For example, when discussing the appropriate application of limitations and restrictions, E1 suggested making decisions based on the needs of each child: *"If I knew a child I would make different decisions according to the child."* (E1). E6 also made a similar recommendation:

*"You have to work at the pace of the child because for some children they might stick at only having five symbols (programming blocks) for quite a long time whereas another child might get it straight away and the next week they want to be writing a story. So you might have to make more available for them more quickly."* (E6)

Personalisation can also be applied in prioritising the child's interest when providing content to create programs with. E7 believed that this can help increase engagement:

*"I think if it's following their interest a bit more I think that's going to make it more interesting to them, isn't it? It's picking up on something they know about and want to learn about. I think that's another additional problem with probably all of them, is finding the things that float their boat, which is a massive issue, which is why you know personalizing it is probably going to help with that difficulty."* (E7)

For personalisation, a VPT would require a knowledge of the characteristics of the child using it, and make decisions based on the characteristics and needs of the child. Therefore, the recommendations proposed here are to have user profiles for storing user information and then to personalise based on the contents of the profile.

*2.4.9. Managing Changes*

Most of the recommendations discussed in the themes above propose some form of modification to the user interface or logic of VPTs to improve accessibility. Implementing these modifications will undoubtedly produce changes that will be noticeable by the user. Although children with ASC are known for their difficulties with dealing with changes, experts suggested informing them about modifications to VPTs before they occur as a way of reducing the difficulty:

*"A characteristic of people with autism is that change can be something that's really challenging but if they know it's going to come and the connotations of the change are good ones then they can find it easy to cope with it."* (E4)

*"Some children will hate it if you change it from the way it was originally but actually in my experience you can, you can warn them, you can explain why, you can present it as a good thing. So, I wouldn't say that it would be wrong to do it, but I think it's how you present it"* (E1)

It can be seen from the excerpts above that although children with ASC characteristically do not respond well to unexpected changes, they deal well with changes that they are made aware of, and understand the benefit of. Therefore making changes slowly and not drastically can make it easier for children with ASC to accept the changes as explained by E2:

*"If you can scaffold it and gently move things from one to two to three to four, you know, be clear about where they are in the journey, I think that the expectations can be managed well enough that it shouldn't represent a major problem."* (E2)

The recommendations proposed for helping children with ASC handle changes made to VPTs as a result of personalisation are: to present notifications whenever a change is due to be made; to avoid making major changes at an instance; and to have a feature that keeps track of changes applied for each child and presents it in a visual manner when requested.

Table 2: Recommendations gathered from analysing interviews.

| Theme | Recommendation | Description |
|---|---|---|
| Mobile Device Compatibility | Make sure VPT is compatible with mobile devices | VPTs should be compatible with and accessible on mobile devices, especially smartphones and tablets in order to allow easier access and interactions for children with ASC including those with motor difficulties. |

Table 2: Recommendations gathered from analysing interviews.

| Theme | Recommendation | Description |
| --- | --- | --- |
| Engaging Users | Integrate content known to interest children with ASC | Provide users with a diverse set of media and templates covering as many topics known to interest children with ASC as possible. For example, provide templates and media related to various forms of transportation e.g. a space rocket launch project template, sprites and models of planets, astronauts etc. |
| | Suggest popular content to unmotivated users. | Content suggestions should be made to users (especially those barely interacting with the VPT) by suggesting content that is popular among other users with similar profiles. |
| Information Presentation and Visualisation | Represent information visually using icons/symbols | Information within VPTs should be represented or supported with visual symbols/icons. For example, visual symbols or icons should be used to label objects or to support the objects' text labels throughout the user interface. This should include labels on buttons, tabs, panes, programming elements (blocks, bricks) etc. |
| Sounds | Sounds should be optional | The VPT should be usable with or without sounds. Sounds (including feedback sounds and program sounds) should not be audible unless explicitly turned on by the user, and volume control should be provided for users to adjust their sound level. |
| Restrictions and Limitations | Limit the choices of media elements available to users to a relevant and manageable set | Users should be provided with a manageable and relevant subset of media items (e.g. characters, backgrounds etc.) to work with, based on the goal of their project. For example, when creating a "space racing game", the choice of background images or landscapes options can be limited to only those related to outer space. |
| | Limit the choices of programming elements available to users to a usable, manageable and relevant set | Provide users with a usable, manageable and relevant subset of programming elements (blocks, bricks etc.) for their projects. For example, only a small subset of basic programming blocks should be available to a user with ASC and severe learning disabilities and ensuring that only blocks that reposition objects are available when the user is creating a script for moving an object. |

Table 2: Recommendations gathered from analysing interviews.

| Theme | Recommendation | Description |
| --- | --- | --- |
| | Limit the features available to the user to those required to achieve the user's goal and are within the user's cognitive abilities | VPTs usually have numerous tools and features for performing various tasks. Only those features that the user is capable of using and support the achievement of the user's goal should be available to the user. For example, a user that requires constant scaffolding should be restricted from using the 'create blank/empty project' feature. Features that become subjects of obsessive behaviour to the extent that they stand in the way of goal achievement should also be restricted to allow the user to move on to achieve their goal. |
| Scaffolding | Provide templates for projects | VPTs should provide templates for creating a wide range of projects (e.g. games and animated stories). The templates should provide a visual structural scaffold for users by guiding them through the various stages of creating a project. For example, templates for a story should guide users to choose characters, backgrounds, and actions for each character. They should also allow users to view a potential version of their end product. |
| | Provide programming elements at higher levels of abstraction | VPTs should provide programming elements at different levels of abstractions for users with different abilities. For example, visual programming elements for 'move along x' and 'move along y' can exist at the low-level. However, for those users that may be unable to create a script using these two programming elements to represent jumping, high-level programming elements 'jump forwards' and 'jump backwards' should be made available. |
| Goal Orientation | Design to ease, support and encourage success and goal achievement | Personalisation, restrictions, limitations, visualisations and scaffolding, should all be applied in a way that helps and encourages users to achieve their goal(s) without having to perform unnecessary/inaccessible actions or tasks. |
| Personalisation | Use profiles to store information about users | VPTs should have user profiles for storing relevant personal information about users (e.g. interests, capabilities and difficulties) and the users' interaction history (e.g. frequency of programming blocks' usage, properties of programs created). User profiles should be automatically updated with each use of the tool. |
| | Personalise based on user profile and interaction history | VPTs should configure their user interface, apply the right restrictions, limitations, choose the right level of programming abstraction etc. based on users' profiles. As the user's profile evolves, the tool should also reconfigure itself to keep up with the changing needs of the user. |

Table 2: Recommendations gathered from analysing interviews.

| Theme | Recommendation | Description |
|---|---|---|
| Managing Changes | Notify users before making any changes | Users should be made aware of any change or changes due to personalisation that will affect the way the VPT looks or functions before said changes are made. The notifications should be subtle and simple to comprehend. |
| | Keep track of changes applied for a user | The VPT should keep a history of changes made to a user's configuration. This should be available for a user to view visually as a form of journey tracker. |

## 3. Validating Recommendations I: Expert Interviews

Some of the initial recommendations gathered were explicitly stated by experts, while others were proposed by the authors to address concerns raised by experts. Therefore, a validation exercise was conducted to confirm the validity of all proposed recommendations. The validation study and its findings are presented in this section.

### 3.1. Participants

The same approach used to recruit participants in the initial study was used for participant recruitment in this study. Three experts, all of whom had participated in the initial study (E1, E4 and E6), agreed to participate.

### 3.2. Method

A qualitative approach to validation was employed using semi-structured interviews as the data collection method due to the success of this approach in the initial study and the richness of the data gathered. Since ethical approval was already secured for the research before the start of the initial study, no new ethics permission was requested.

All interviews took place in May 2019, were held face to face, lasted between 30 – 45 minutes and were audio-recorded with the permission of the interviewee. Before the start of each interview, each expert was given a copy of the proposed recommendations. Then a semi-structured interview approach was adopted to investigate the validity of the recommendations previously derived within each theme. Based on the expert's response, the interviewer then followed up on any interesting points raised to either validate or correct existing recommendations, or even to derive additional recommendations.

### 3.3. Analysis

Interesting concepts were first identified by listening to the interviews. Then the interviews were transcribed and coded. Coding was done by examining the data for opinions, suggested modifications, additions or eliminations. When any of these were identified, it was coded and assigned to the theme it relates to (no code was found to require a new theme). Then the codes in each theme followed a similar process as specified above. Findings showed agreement on the validity of all recommendations, two recommendation additions, no eliminations, and some suggested modifications.

QSR Nvivo 12 was used to conduct this data analysis. It was used to transcribe audio interview data, store notes, code data, and update themes.

### 3.4. Findings

The findings of the validation exercise are discussed in this section. Excerpts from the interview transcripts are used to show experts' opinions about the recommendations in their own words. The changes made to the gathered recommendations as a result of the findings are presented in Table 3. Newly added recommendations and their descriptions, and updated recommendations for initial recommendations are presented in italics.

#### 3.4.1. Mobile Device Compatibility

There was a general agreement by experts on the need for VPTs to be available on mobile devices:

*"It would be fantastic if it could be very portable, for children who can engage on phones that would be great. I think as a minimum it should be on some kind of tablet."* (E1)

*"If you can give them the opportunity to create something where they haven't got to be holding a pen or a mouse and they can create an animation or whatever it's going to be, I think that would be really important because we don't want to set up an alternative means of doing something that still has challenges for them. I think most children understand drag and drop because they do it all the time in games and things."* (E6)

### 3.4.2. Engaging Users

Experts welcomed the recommendation for integrating content that is known to interest children with ASC within VPTs. E6 also suggested making the most out of this type of content to capture the attention of users:

*"Make sure that some of their favourite things are on the first screen that they see, I think that would be really important. If it's sharks or rainbows or unicorns you know, whatever their thing is 'my favourite thing is going to be behind the screen, so I need to keep going'"* (E6)

However, with regards to making popular content suggestions to unmotivated users, E4 warned that this might not be suitable for those that have highly restrictive interests:

*"It may well be that in their profile you get a question about their resistance to new information. I can imagine scenarios where if you're really into Thomas the Tank Engine and somebody keeps saying you should use my little pony, then it won't go down very well."* (E4)

Considering these findings, the descriptions for the recommendations related to the integration of engaging content were modified.

### 3.4.3. Information Presentation and Visualisation

Although all experts agreed on the need for visualisation using icons/symbols, concerns over the choice of using new or existing symbols/icons sets were raised:

*"Regardless of what they've been used to, some children will be able very quickly to pick up a new [visual] language, a new set of icons, they'll just quickly figure it out. And others obviously might want it to be something more familiar."* (E1)

E4 believed that using existing sets will be best since children with ASC will already be familiar with them, and if they are not, then they get the chance to learn a symbol set that will be useful to them in other contexts. This also avoids putting the children through the unnecessary task of learning to understand the visual language in order to be able to use the tool. However, the use of an existing symbol set might come with certain Intellectual Property and Rights restrictions, and different children might be experienced in the use of different sets, which may mean having VPTs support multiple sets and to record individual preferences within profiles. Therefore, E6 favoured the idea of having new sets created and used for VPTs, and argued that as long as the children are interested, they will be able to learn the new symbol set:

*"I found when we have introduced…, whether it's for communication devices or programming, you know when kids started getting into the Lego programming and model making things because they're engaged with it and they are enthused, they can quite quickly learn the rules and the symbols that go with it. So I think if you have a new set and they're excited to use it, they will be able to learn"* (E6)

Considering these findings, we propose the use of existing visual/iconic languages where possible and only use new ones in cases where the use of existing ones is not possible due financial, legal or other constraints.

### 3.4.4. Sounds

The need for making all sounds optional for the use of VPTs was also agreed on by experts.

### 3.4.5. Restrictions and Limitations

Experts all agreed on the need to put restrictions and limitations to ensure users are focused and not overwhelmed:

*"I think it's really important because when you give too many choices, it might look great, but they just don't know where to begin. It's linked to their executive function challenges, being able to differentiate between what's important and not important, and how to organize things"* (E6)

However, when imposing restrictions to help users get out of repetitive cycles, E4 advised against imposing them on content used to create programs even if the user is fixated on that content:

*"I think if they are only fixated on some topics then you should exploit that and remember that you are trying to get them to move up the levels of coding."* (E4)

The corresponding recommendation was updated to reflect the additional information gathered.

### 3.4.6. Scaffolding

All experts agreed on the need for structure using templates and the provision of scaffolds, to ensure children with ASC with different needs and abilities can program using VPTs:

*"It's an opportunity to really engage in a different way with something they've never done and it might…, it should be highly motivating especially if it scaffolds at the right level"* (E1)

*"The use of templates is a good idea, particularly for the children having moderate or severe learning difficulties"* (E6)

### 3.4.7. Goal Orientation

This recommendation was also well received by experts. E6 mentions that this design approach can also serve as a way of giving caregivers ideas on how VPTs can be used by children with ASC, thus encouraging them to introduce the use of the tools at home or in class:

*"I mean people pick out particular programs and tools because they think it will help with a particular goal. So this would be another way of getting the caregivers interested in picking this software."* (E6)

E6's insight led to the addition of a new recommendation for ensuring VPTs are goal oriented.

### 3.4.8. Personalisation

Recommendations concerning personalisation were greatly supported by all experts. However, a concern was raised by both E6 and E1 regarding the time and precision required to accurately set up profiles:

*"The user profile presumably is something that a teacher or parent or someone else might be able to put in as opposed to the child themselves right? So all of these things will take a certain amount of setting up"* (E6)

To tackle this issue and to improve the accuracy of the information within profiles, E1 suggested suing a user-modelling test for children with ASC to automatically generate their profiles. Additionally, E1 suggested the provision of a feature that allows caretakers or the users themselves to make customise VPTs in addition to the automatic personalisations. This is useful especially in cases where the applied automatic personalisation is not producing the expected positive results.

### 3.4.9. Managing Changes

Experts all agreed that these recommendations are necessary for managing the frustrations that children with ASC face when dealing with changes.

Table 3: Changes to recommendations as a result of validation

| Theme | Recommendation | Description |
| --- | --- | --- |
| Engaging Users | Integrate content known to interest those with ASC | In addition to the initial description: *Where possible, make use of such media items on splash screens, lock screens etc. to capture the attention of users.* |
| | Suggest popular content to unmotivated users | In addition to the initial description: *This should only be applied for users that are not resistant to new information.* |
| Information Presentation and Visualisation | Represent information visually using icons/symbols | In addition to the initial description: *Restrictions should focus on features and not media content or programming elements, even if the user is fixated on them.* |
| Goal Orientation | *Provided templates should scaffold towards projects appropriate for teaching relevant skills to children with ASC.* | *Templates should be designed for projects that teach children with ASC relevant skills such as communication and collaboration.* |
| Personalisation and Customisation | *Store personal user information and preferences.* | *VPTs should have user profiles for storing relevant personal information about users (e.g. interests, capabilities and difficulties). An automated user modelling test can be used to collect user data for initialising the user's profile, otherwise, the data can be entered manually by a caretaker.* |
| | *Record users' interaction history for personalisation.* | *VPTs should record the users' use of the tool (e.g. frequency of programming blocks' usage, properties of programs created). This record should be automatically updated with each use of the tool.* |
| | *Support user customisation, i.e. manual selection of preferences.* | *The VPT should also allow manual setting of preferences as a way of overriding automatic personalisation. For example a user should be able to choose font size and colour.* |

## 4. Validating Recommendations II: Examination of Related Interactive Applications Literature to Identify Remaining Gaps

This section presents further validation of the proposed recommendations for designing VPTs for children with ASC that have resulted from our study by comparing with those for other interactive applications for the same target group. This process will serve as a final validation for our derived guidelines, and identify any remaining gaps.

Scopus was queried to identify peer-reviewed research works that provide recommendations for designing interactive applications for children with ASC. Keywords used for the query include ASC, Autism, ASD, Software, Games, Applications, VR, Recommendations and Guidelines. The search was limited to provide results that were published from 2010 to 2019. Seven relevant research papers were identified from the results of this search. Google Scholar was also queried using the same keywords that were used to query Scopus. Three additional research papers were identified from the results of Google Scholar.

Although none of the recommendations found in the literature are targeted for designing VPTs for those with ASC, recommendations were found for designing serious games (Tsikinas and Xinogalos, 2019), websites (Britto and Pizzolato, 2016; Raymaker et al., 2019), mobile applications (Dattolo and Luccio, 2017), VR applications (Bozgeyikli et al., 2018; Herrera et al., 2018), tangible user interfaces (Sitdhisanguan et al., 2012) and other interactive applications (Davis et al., 2010; Pavlov, 2014; Khowaja and Salim, 2015). Various methodologies were used in gathering these guidelines including literature reviews (Bozgeyikli et al., 2018; Britto and Pizzolato, 2016; Dattolo and Luccio, 2017; Tsikinas and Xinogalos, 2019; Herrera et al., 2018), engaging stakeholders Raymaker et al. (2019), extending existing guidelines (Khowaja and Salim, 2015), combination of literature review and stakeholder engagement (Pavlov, 2014), and from research experiences (Davis et al., 2010; Sitdhisanguan et al., 2012).

### 4.1. Comparing Initial Recommendations with Recommendations from the Literature

Each set of recommendations found in the literature was analysed to identify similarities to those gathered and validated by this study. Recommendations from the literature that fit the themes identified in this study were first highlighted, then those similar to individual recommendations gathered by this study were identified.

Dattolo and Luccio (2017) proposed recommendations for designing websites and mobile applications for those with ASC and made strong arguments for the suitability of mobile apps for children with ASC. Britto and Pizzolato (2016) did not recommend the support for mobile devices but they recommended having the appropriate sensitivity on touch screens to prevent selection and accidental touch errors.

As recommended by this study, Bozgeyikli et al. (2018) also recommend taking advantage of the special interests of children with ASC to provide engaging and motivating content in virtual reality applications. Similarly, Davis et al. (2010) recommended accommodating special interests of children with ASC when designing interactive applications, they also recommended avoiding content with any fears the user may have. Another way of engaging children with ASC that lose interest and do not perform any interactions for a while is to gain their attention using a relevant stimulus e.g. sound or visual cue (Sitdhisanguan et al., 2012).

Visual presentation of information was found to be a popular recommendation in the literature. Britto and Pizzolato (2016); Dattolo and Luccio (2017); Pavlov (2014); Raymaker et al. (2019) all recommend having visual objects as an alternative means of presenting information. When using icons or symbols, Britto and Pizzolato (2016); Khowaja and Salim (2015); Raymaker et al. (2019) recommend using those representing concrete actions that can easily be recognised by users. Wherever text is used, recommendations suggest the use of an accessible font-type (Pavlov, 2014; Tsikinas and Xinogalos, 2019; Raymaker et al., 2019; Britto and Pizzolato, 2016), and using simple straightforward language with no jargon, acronyms etc. (Khowaja and Salim, 2015; Britto and Pizzolato, 2016; Raymaker et al., 2019; Dattolo and Luccio, 2017).

Although Bozgeyikli et al. (2018) note that some studies have found positive impacts in the use of sounds to improve user motivation in VR applications, they still recommend making sounds optional. Dattolo and Luccio (2017); Davis et al. (2010); Pavlov (2014) do not recommend making sounds optional but they do recommend avoiding loud and unnecessary sounds.

Applying restrictions and limitations on VPTs is part of the recommendations proposed in this study, including among other things, improving goal achievement and reducing repetitive tendencies. Bozgeyikli et al. (2018); Davis et al. (2010) both recommend taking over control at certain times by designing out or preventing the user from performing certain tasks or accessing certain features for the same reasons.

Scaffolding experiences by providing structuring templates and highly abstracted programming elements may be a useful recommendation for designing VPTs, but not necessarily for other applications such as games, VR applications and websites. Thus other scaffolding approaches were found in the literature including providing multimedia instructions for interacting with interface objects (Britto and Pizzolato, 2016; Pavlov, 2014), and providing relevant examples (Raymaker et al., 2019).

Integrating personalisation and customisation capabilities when designing for children with ASC is highly supported by the recommendations found in the literature. Aspects that are recommended for personalisation and customisation include text size, colour and font (Britto and Pizzolato, 2016; Khowaja and Salim, 2015; Pavlov, 2014), characters and environments (Bozgeyikli et al., 2018; Tsikinas and Xinogalos, 2019), and number of elements within the interface (Britto and Pizzolato, 2016).

In ensuring that children with ASC are focused on their goals, recommendations have been made to avoid displaying distracting elements on-screen (Britto and Pizzolato, 2016; Dattolo and Luccio, 2017; Khowaja and Salim, 2015;

Pavlov, 2014) and to provide access only to features that help in goal achievement (Britto and Pizzolato, 2016; Davis et al., 2010; Khowaja and Salim, 2015). Both recommendations are in line with the recommendation proposed by this study in ensuring design encourages and supports goal achievement.

Only Khowaja and Salim (2015) made a recommendation that addressed the need to handle changes to applications in a step by step approach since children with ASC do not cope well with drastic changes. Another similar recommendation, not restricted to interface changes, proposed keeping users informed about the status of the system and providing constant feedback to users (Britto and Pizzolato, 2016).

### 4.2. Relevant Recommendations from the Literature

In addition to the recommendations from the literature that align with the themes identified by our study, several other recommendations found in the literature relating to other interactive applications could be applied to the design of VPTs. They have been extracted, categorised into themes and presented below.

#### 4.2.1. User Interface and Navigation

User interfaces should be designed with simple structures, predictable and no distracting secondary content (Pavlov, 2014; Khowaja and Salim, 2015; Davis et al., 2010; Raymaker et al., 2019; Dattolo and Luccio, 2017; Tsikinas and Xinogalos, 2019; Bozgeyikli et al., 2018). Mild colours should be used and bright colours avoided (Pavlov, 2014; Bozgeyikli et al., 2018), and there should be a clear differentiation between background and foreground elements (Pavlov, 2014; Britto and Pizzolato, 2016; Bozgeyikli et al., 2018; Sitdhisanguan et al., 2012). Icons and buttons should be big enough to be clickable and look clickable (Pavlov, 2014; Britto and Pizzolato, 2016). User interfaces should also include consistent navigation, with no automatic redirects or time limit before a page should be exited (Pavlov, 2014; Britto and Pizzolato, 2016; Raymaker et al., 2019; Dattolo and Luccio, 2017).

#### 4.2.2. System Status

Visual indicators should be used to inform users about the duration or waiting period associated with any time-consuming actions (Pavlov, 2014).

#### 4.2.3. Control

Actions should be easily cancelled, reverted, undone or confirmed in order to resolve errors quickly and to encourage exploration without the fear of consequences (Khowaja and Salim, 2015; Britto and Pizzolato, 2016).

#### 4.2.4. Low Latency

All actions should be handled quickly and feedback provided. Latency should be avoided as it can easily frustrate children with ASC (Khowaja and Salim, 2015).

#### 4.2.5. Accessible Documentation

Documentation designed with children with ASC in mind should be easily retrievable and accessible at any time to provide relevant multimedia help aimed at helping them complete their current task (Khowaja and Salim, 2015).

## 5. Discussion

The aim of this study is to propose recommendations for designing accessible, usable, and engaging VPTs for children with ASC. A literature review was carried out but no recommendations were found for designing accessible and usable VPTS for children with ASC and cognitive impairments. In response to this identified gap, the first part of the current study gathered recommendations by conducting semi-structured interviews with seven experts and using thematic analysis to analyse the interviews. This led to the proposal of 16 recommendations categorised under nine themes identified from the findings. Although all the themes represented areas of concerns highlighted by experts, not all the recommendations were proposed by the experts themselves, some were proposed by the authors to mitigate issues raised by the experts. To ensure the validity both recommendations by the experts and the authors, a validation exercise was conducted by using semi-structured interviews with three experts. Findings from this validation exercise ensured the validity of all the gathered recommendations, and also provided new data that led to the addition of two new recommendations. Several initial recommendations were also updated to further clarify their meaning. This was mainly done by rewording or adding content to their descriptions, although in one case one recommendation was broken into two separate recommendations. This increased the total number of proposed recommendations to 19, while the number of themes remained unchanged.

Finally, another validation step was taken to ensure the gathered and validated recommendations are in line with recommendations for designing other non-VPT related interactive applications proposed in the literature, and to identify any gaps. The recommendations found in the literature supported all the identified recommendation themes, but not all recommendations within the themes. However, 5 out of the 7 recommendations that are not supported by the literature are specific to VPTs e .g. the recommendation that project templates should be p rovided. Of the gathered and validated recommendations that are supported by the literature, 4 are supported by only 1 set of recommendations from the literature, 3 are supported by 2 sets of recommendations from the literature, 2 are supported by 4 sets recommendations from the literature, 1 is supported by 5 sets recommendations from the literature, 1 is supported by 6 sets recommendations from the literature, and a final 1 supported by all 10 sets recommendations found in the literature. Recommendations supported by a higher of supporting recommendations does not indicate a greater degree of confidence in the derived r ecommendation. It indicates that this recommendation is applicable to a higher number of domains.

In addition to validating our recommendations, the examination of the literature concerning non-VPT related interactive applications also exposed gaps that were missed during interviews with experts, and hence missed by our proposed recommendations. It should be noted that some recommendations found in the literature appeared to contradict the ones gathered by this study. This is because these recommendations are specifically suitable for the type of interactive application they are being recommended for but not for VPTs. An example is the recommendation by Tsikinas and Xinogalos (2019) to support the repetition of tasks within serious games, so that children with ASC can enjoy this repetition and also master the task. While this makes sense in the context of serious games which are used to master a particular skill, it does not fit VPTs well since they are used to encourage learning through exploration and creation of interactive media products. Therefore intentionally supporting repetition will defeat the purpose of using VPTs within this learning context. A more suitable recommendation is the one gathered from experts, which suggests restricting the features available to children with ASC, and allowing them to explore and create within a more structured environment but still with some degree of freedom depending on their needs.

There were also recommendations from the literature that were suitable for designing VPTs for children with ASC but were not covered by our gathered recommendations. The methodology in the original interviews framed the discussion around a previous study that investigated the accessibility of Scratch 2.0. It could have been possible that this imposed some limit on the responses provided by the experts and hence the recommendations derived from those interviews. However, we feel this unlikely given the breadth of the experts' experience in the design of many technological based interventions for children with ASC and associative cognitive impairments. But we remain open to these additionally derived recommendations and include them in our final s et. Each of these additional recommendations was either added into an appropriate existing theme or into a newly created theme if an appropriate one does not exist. Seven new recommendations were added into existing themes, and 9 new recommendations were added into 5 new themes.

The complete set of proposed recommendations are provided in Table 4 (the final recommendations informed by the literature relating to non-VPT interactive applications and not identified in the interviews with experts are presented in italics). They are presented in a descriptive manner intentionally to ensure their applicability to VPTs of different nature (e.g. those that support block based programming or those that have a game based programming environment), and to allow innovation in the way they are applied by designers and developers. However to avoid ambiguity, concrete examples have been provided (where it seems necessary) to show how the recommendations can be applied to existing VPTs, however, these examples are in no way meant to be exhaustive.

Table 4: Complete set of proposed recommendations

| Theme | Recommendation | Description |
|---|---|---|
| Mobile Device Compatibility | Make sure VPT is compatible with mobile devices | VPTs should be compatible with and accessible on mobile devices, especially smartphones and tablets in order to allow easier access and interactions for children with ASC including those with motor difficulties. |
| | *Touch screen interactions should have the appropriate sensibility and prevent errors in selections and accidental touch in interface elements* | *VPTs should be designed to support use by children with fine motor skill difficulties by having appropriate sensibility and preventing errors associated with the condition.* |
| Engaging Users | Integrate content known to interest children with ASC | Provide users with a diverse set of media and templates covering as many topics known to interest children with ASC as possible. For example, provide templates and media related to various forms of transportation e.g. a space rocket launch project template, sprites and models of planets, astronauts etc. Where possible, make use of such media items on splash screens, lock screens etc. to capture the attention of users. |
| | Suggest popular content to unmotivated users | Content suggestions should be made to users (especially those barely interacting with the VPT) by suggesting content that is popular among other users with similar profiles. This should only be applied for users that are not resistant to new information. |
| | *Stimulate users after a period of inactivity* | *VPTs should be able to attract the attention of users that lose interest and do not perform any interactions for a while using a relevant stimulus e.g. sound or visual cue, this may vary depending on the preference and interest of the child.* |
| Information Presentation and Visualisation | Represent information visually using icons/symbols | Information within VPTs should be represented or supported with visual symbols/icons. For example, visual symbols or icons should be used to label objects or to support the objects' text labels throughout the user interface. This should include labels on buttons, tabs, panes, programming elements (blocks, bricks) etc. Existing symbol/icon sets should be used where possible, otherwise a new, easy to understand set of symbols/icons can be created and used. |
| | *Use clear accessible font for text* | *VPTs should support the use of fonts that are accessible and easy to read by thgose with ASC e.g. Arial.* |

Table 4: Complete set of proposed recommendations

| Theme | Recommendation | Description |
|---|---|---|
| | *Language used should be simple, consistent, precise, with no jargon and with concepts and phrases familiar to users* | *VPTs should support language that is easily understandable by children with ASC, and avoid using technical jargon, abbreviations etc.* |
| Sounds | Make sounds optional, turn it off by default and allow users to control volume | The VPT should be usable with or without sounds. Sounds (including feedback sounds and program sounds) should not be audible unless explicitly turned on by the user, and volume control should be provided for users to adjust their sound level. |
| | *Avoid disturbing and explosive sounds or any other unnecessary sounds* | *Any loud or disturbing sound should be avoided, especially as feedback sounds. Explosions, sirens etc, should only be played when chosen by the user.* |
| Restrictions and Limitations | Limit the choices of media elements available to users to a reasonable and meaningful set | Users should be provided with a manageable and relevant subset of media items (e.g. characters, backgrounds etc.) to work with, based on the goal of their project. For example, when creating a "space racing game", the choice of background images or landscapes options can be limited to only those related to outer space. |
| | Limit the choices of programming elements available to users to a reasonable and meaningful set | Provide users with a usable, manageable and relevant subset of programming elements (blocks, bricks etc.) for their projects. For example, only a small subset of basic programming blocks should be available to a user with ASC and severe learning disabilities and ensuring that only blocks that reposition objects are available when the user is creating a script for moving an object. |
| | Limit the features available on the interface to the ones that are needed to achive the user's goal, and restrict access to features that control redundant things | VPTs usually have numerous tools and features for performing various tasks. Only those features that the user is capable of using and support the achievement of the user's goal should be available to the user. For example, a user that requires constant scaffolding should be restricted from using the 'create blank/empty project' feature. Features that become subjects of obsessive behaviour to the extent that they stand in the way of goal achievement should also be restricted to allow the user to move on to achieve their goal. Restrictions should focus on features and not media content or programming elements, even if the user is fixated on them. |

Table 4: Complete set of proposed recommendations

| Theme | Recommendation | Description |
| --- | --- | --- |
| Scaffolding | Provide templates for projects | VPTs should provide templates for creating a wide range of projects (e.g. games and animated stories). The templates should provide a visual structural scaffold for users by guiding them through the various stages of creating a project. For example a template for a story should guide users to choose characters, backgrounds, and actions for each character. They should also allow users to view a potential version of their end product. |
| | Provide programming elements at higher levels of abstraction | VPTs should provide programming elements at different levels of abstractions for users with different abilities. For example, visual programming elements for 'move along x' and 'move along y' can exist at the low-level. However, for those users that may be unable to create a script using these two programming elements to represent jumping, high-level programming elements 'jump forwards' and 'jump backwards' should be made available. |
| | *Provide concrete examples where applicable, to accommodate difficulties in understanding concepts* | *Provide a library of examples of projects and how concepts are used for users to learn from.* |
| | Present appropriate instructions to interact with interface elements | Instructions should be available to guide users on how to use the various interface elements available in VPTs, ideally this should also be visual and accessible for children with ASC. |
| Personalisation and Customisation | Store personal user information and preferences | VPTs should have user profiles for storing relevant personal information about users (e.g. interests, capabilities and difficulties). An automated user modelling test can be used to collect user data for initialising the user's profile, otherwise, the data can be entered manually by a caretaker. |
| | Record users interaction history | VPTs should record the users' use of the tool (e.g. frequency of programming blocks' usage, properties of programs created). This record should be automatically updated with each use of the tool. |
| | Personalise based on user profile and interaction history | VPTs should configure their user interface, apply the right restrictions, limitations, choose the right level of programming abstraction etc. based on users' profiles. As a user's profile evolves, the tool should also reconfigure itself to keep up with the changing needs of that user. |

Table 4: Complete set of proposed recommendations

| Theme | Recommendation | Description |
|---|---|---|
| | Support user customisation, i.e. manual selection of preferences | The VPT should also allow manual setting of preferences as a way of overriding automatic personalisation. For example a user should be able to choose font, size and colour. |
| Goal Orientation | Design to ease, support and encourage success and goal achievement | Personalisation, restrictions, limitations, visualisations and scaffolding, should all be applied in a way that helps and encourages users to achieve their goal(s) without having to perform unnecessary/inaccessible actions or tasks. |
| | Provided templates should scaffold towards projects appropriate for teaching relevant skills to children with ASC | Templates should be designed for projects that teach children with ASC relevant skills such as communication and collaboration. |
| Managing Changes | Notify users before making any changes | Users should be made aware of any change or changes due to personalisation that will affect the way the VPT looks or functions before said changes are made. The notifications should be subtle and simple to comprehend. |
| | Implement changes in small and manageable steps | Changes should be made in small manageable steps that can be handled by users. Drastic changes with major impacts should not be implemented at once. |
| | Keep track of changes applied for a user | The VPT should keep a history of changes made to a user's configuration. This should be available for a user to view visually as a form of journey tracker. |
| *User Interface and Navigation* | *The user should be able to control navigation and time to perform a task* | *VPTs should avoid automatic redirects and allow users to have total control over navigation. No time limit should be used to determine how long a user stays in a section of a VPT.* |
| | *Always differentiate between background colour and foreground objects* | *Contrasting colours should be used to differentiate between the background and objects in the foreground. For example colours of programming elements should be used to differentiate them from their background.* |
| | *Use mild colours and avoid bright colours* | *Bright colours should be avoided to ensure that the sensitivities experienced by children with ASC connected to bright colours are not triggered.* |
| | *The design and structure should be simple, clear and predictable* | *The overall design of VPTs should be easy to learn to use, navigate and should not contain any surprises. It should be predictable and consistent in all aspects.* |

Table 4: Complete set of proposed recommendations

| Theme | Recommendation | Description |
|---|---|---|
| | *Clickable icons, buttons and other interactive elements should be big enough to provide appropriate click/tap area* | *Clickable icons, buttons etc. in VPTS should be designed to appear clickable to the users, and they should be big enough to be easily clicked or tapped by users even those with fine motor skills difficulties.* |
| Low Latency | Avoid the frustration of users by avoiding latency of supported actions | Children with ASC can be easily distracted or frustrated during periods of inactivity, or while waiting for actions to be complete. Therefore VPTs designed for them should be quick in executing actions and providing feedback. |
| System Status | Visual indicators should be used to inform users about time consuming actions | VPTs may perform actions that take some time to be completed e.g. compiling a program before executing or downloading graphic assets. Whenever these actions are performed visual indicators should be used to inform the user about their corresponding status. |
| Control | Allow critical actions to be reverted, cancelled, undone or confirmed | To avoid frustrations and to encourage exploration and creativity, VPTs should allow children with ASC to easily cancel, revert or undo their actions. |
| Accessible Documentation | Documentation, help or instructions should be visible | VPTs' documentation should be visible to users, or it should be easily retrievable when needed, it should focus on the user's task, and provide multimedia demonstration of tasks. |

## 6. Limitations

Like any research investigation, this study has its limitations. Although all experts that were interviewed had extensive experience in designing interventions for children with ASC (either traditional or technology-based), not all were familiar with VPTs before participating in this study. To address this, all experts were shown an example of a VPT and a demonstration of how it can be used to create a simple animation before the interview started.

Secondly, in the first part of the study, the interview was mainly driven by discussions around the difficulties and behaviours faced by five children with ASC observed in a related study that evaluated the accessibility of Scratch 2.0 for children with cognitive impairments. However, we believe the recommendations that were gathered as a result of discussing these observations can be generalised to making all VPTs accessible.

Lastly, the final set of recommendations proposed by this study contains those validated by experts and unvaldiated recommendations extracted from the literature. The selection of the latter set of recommendations informed by the literature is based on the authors previous experience in assessing the accessibility of specific VPT i.e. Scratch 2.0. Although these recommendations are based on well documented recommendations for designing various non-VPT interactive applications, they have not as yet been validated by the experts particpating in this study. Therefore we offer them as tentative recommendations for the design of VPTs for children with ASC and are italised in Table 4.

Very often, when deriving guidelines on the subject of accessibility for specific target end users, the guidelines themselves are not accessible to these very end users. To address this concern we have started to produce an accessible version of our accessibility recommendations for designing VPTs using symbols, images and simplified text.

# 7. Conclusion and Future Work

In spite of the popularity of VPTs for use by neurotypical children, and the reported benefits of using them in learning contexts, there is still a lack of research on their use for children with ASC, especially those with a learning disability. Following on from previous research on evaluating the usability of VPTs for children cognitive impairments (Zubair et al., 2018), including those with ASC, this work aimed to propose a set of recommendations for designing accessible, usable and engaging VPTs for children with ASC. An initial set of recommendations were gathered from interviewing seven experts before they were updated after validation by interviewing a subset of the previous seven experts (three in total). On comparison with recommendations for designing other non-VPT interactive applications for those with ASC found in the literature, the validity of the gathered recommendations was further confirmed, and additional relevant recommendations were derived. This comparison also showed that although there are accessibility recommendations that can be universally applied to all types of interactive applications, some are specific to certain types of applications and domains (e.g. the recommendation to support repetition in serious games).

As future work, the recommendations proposed within the personalisation theme will be implemented into an existing VPT. The implemented personalisation will mainly be applied to scaffolding and limitations related to programming elements. An ontology will be used to represent programming elements categorised based on complexity and purpose, and how they should be made available to users depending on their characteristics (as represented in their profiles). Three personae created for children with ASC by the authors will be used to create the user profiles that will be used in the ontology.

# Acknowledgement

This study was funded in part by Petroleum Technology Development Fund (PTDF).